\documentstyle{article}
\def\bea{\begin{eqnarray}}
\def\eea{\end{eqnarray}}
\def\nn{\nonumber}
\def\beq{\begin{equation}}
\def\eeq{\end{equation}}
\def\ba{\beq\new\begin{array}{c}}
\def\ea{\end{array}\eeq}
\def\be{\ba}
\def\ee{\ea}

\def\la{\ \left\langle\hspace{-0.2cm}\left\langle\ }
\def\ra{\ \right\rangle\hspace{-0.2cm}\right\rangle\ }
\def\tm{t^{(m)}}
\def\tn{t^{(n)}}

\makeatletter
\newdimen\normalarrayskip 
\newdimen\minarrayskip 
\normalarrayskip\baselineskip
\minarrayskip\jot
\newif\ifold \oldtrue \def\new{\oldfalse}
\def\arraymode{\ifold\relax\else\displaystyle\fi} 
\def\eqnumphantom{\phantom{(\theequation)}} 
\def\@arrayskip{\ifold\baselineskip\z@\lineskip\z@
\else
\baselineskip\minarrayskip\lineskip2\minarrayskip\fi}
\def\@arrayclassz{\ifcase \@lastchclass \@acolampacol \or
\@ampacol \or \or \or \@addamp \or
\@acolampacol \or \@firstampfalse \@acol \fi
\edef\@preamble{\@preamble
\ifcase \@chnum
\hfil$\relax\arraymode\@sharp$\hfil
\or $\relax\arraymode\@sharp$\hfil
\or \hfil$\relax\arraymode\@sharp$\fi}}
\def\@array[#1]#2{\setbox\@arstrutbox=\hbox{\vrule
height\arraystretch \ht\strutbox
depth\arraystretch \dp\strutbox
width\z@}\@mkpream{#2}\edef\@preamble{\halign
\noexpand\@halignto
\bgroup \tabskip\z@ \@arstrut \@preamble \tabskip\z@ \cr}%
\let\@startpbox\@@startpbox \let\@endpbox\@@endpbox
\if #1t\vtop \else \if#1b\vbox \else \vcenter \fi\fi
\bgroup \let\par\relax
\let\@sharp##\let\protect\relax
\@arrayskip\@preamble}
\def\eqnarray{\stepcounter{equation}%
\let\@currentlabel=\theequation
\global\@eqnswtrue
\global\@eqcnt\z@
\tabskip\@centering
\let\\=\@eqncr
$$%
\halign to \displaywidth\bgroup
\eqnumphantom\@eqnsel\hskip\@centering
$\displaystyle \tabskip\z@ {##}$%
\global\@eqcnt\@ne \hskip 2\arraycolsep
$\displaystyle\arraymode{##}$\hfil
\global\@eqcnt\tw@ \hskip 2\arraycolsep
$\displaystyle\tabskip\z@{##}$\hfil
\tabskip\@centering
&{##}\tabskip\z@\cr}
\begingroup\ifx\undefined\newsymbol \else\def\input#1 {\endgroup}\fi

\begin{document}

\setcounter{footnote}{1}
\def\thefootnote{\fnsymbol{footnote}}
\begin{center}
\hfill FIAN/TD-11/00\\
\hfill ITEP/TH-26/00\\
\hfill hep-th/0005280\\
\vspace{0.3in}
{\Large\bf On Renormalization Group in Abstract QFT
}
\end{center}
\centerline{{\large
A.Mironov}\footnote{Theory Dept., Lebedev Physical Inst. and ITEP, Moscow,
Russia} and {\large A.Morozov}\footnote{ITEP, Moscow, Russia}}

\bigskip

\abstract{\footnotesize  The basics of RG equations for generic
partition functions are briefly reviewed,
keeping in mind an application to the Polyakov-de Boer-Verlindes
description of the holomorphic RG flow.
}

\begin{center}
\rule{5cm}{1pt}
\end{center}

\bigskip
\setcounter{footnote}{0}
\section{RG versus (generalized) Laplace equations}

The notion and properties of the (generalized) renormalization
group (RG) \cite{RG,Pol,RGrev} 
acquire a new attention last years, primarily
because of the interest to its hidden (quasi)integrable
sturcture (related to Whitham dynamics \cite{Kri},
see \cite{GKMMM} and references therein) and to its
implicit occurence in the phenomena like AdS/CFT
correspondence \cite{AdS/CFT} (these studies are focused
on the so called holomorphic RG flows \cite{Hol,Holl,Polyakov,Ver}).
The purpose of this note is to give a concise survey of
the basics of abstract RG theory, separating generic
features from the pecularities of particular models.
One of our goals is to explicitly formulate the
controversies between the different concepts, which
people are now trying to unify. It is a resolution of these
controversies, which should provide a real understanding.

The basic notion of modern quantum field/string theory 
is partition function (exponentiated effective action,
the generating function for all the
correlation functions in the theory in all possible vacua),
resulting from functional integration over fields with an  
action, formed by a complete set of operators.
Importantly, there are two different notions of completeness,
see sect.\ref{comp} below for the definitions.
The exact RG {\it a la} J.Polchinski \cite{Pol}, possessing
a formulation in terms of the diffeomorphisms in the
moduli space ${\cal M}$ of theories \cite{GMS},
requires a {\it strong} completeness. It guarantees
that the linear differential equations emerge for
the partition function, and exponentiation of the
corresponding vector field provides  one-parametric
families (flow lines) in ${\it Diff} {\cal M}$ 
describing the RG flow.
At the same time, in most interesting examples only
a {\it weak} completeness is assumed\footnote{It goes without
saying that in conventional field theories one rarely
requires even a weak completeness. Then the notion of exact
renormalization group is usually
substituted by a renormalizability
requirement, which concerns nothing but the singularities
in the vicinities of the critical points,  ultraviolet 
or infrared.
}: it is the one which is supposedly enough for
integrability etc \cite{UFN2}.
However, as natural for integrable systems, the weak
completeness implies that partition function (interpreted
as an element of some Hilbert space, see \cite{GMS})
satisfies some differential equation in ${\cal M}$,
which are rarely {\it linear} in time (coupling constant) derivatives.
The typical example is the Casimir (generalized Laplace) 
equation for the zonal
spherical functions, if of the second order it is an
ordinary Laplace equation. The other avatars of the same equation
are the $W$- (in particular Virasoro) constraints in matrix models 
\cite{Vir,MM,Wc,Mamorev} and the Hamilton-Jacobi equations for the
large-$N$ Yang-Mills partition functions \cite{Polyakov},
studied in the context of the AdS/CFT correspondence 
\cite{AdS/CFT,Hol,Ver}.
Unfortunately, non-linear differential equations do not possess
an RG-like interpretation, at least naively. They are rather
associated with the huge group ${\it Dop} {\cal M}$
of all differential operators
on ${\cal M}$, and there is no obvious way to associate them 
with the elements of a much smaller diffeomorphism group
${\it Diff} {\cal M}$. 
Perhaps surprisingly, an old result from the matrix models theory
\cite{MMMM} implies that some intermediate notion can exist:
despite matrix model provides only a weakly complete
partition function, an explicit transformation of time-variables
(coupling constants) is known, which (almost) eliminates
the dependence on the size $N$ of the matrix -- what one would
expect the exact RG to provide. A possible explanation
of this phenomenon is that the {\it entire} set of the Ward
identities (i.e. the full power of integrability) was used
in the calculation of \cite{MMMM}, not just a single Virasoro
constraint $L_2$ (which is a counterpart of conventional 
Polchinski's equation).

In refs.\cite{Ver} a seemingly artificial trick was suggested
to resolve the controversy between the linear (in coupling-constants
derivatives) notion of RG and the quadratic (at best)
form of the Lap\-la\-ce/Vi\-ra\-so\-ro/Hamil\-ton-Ja\-co\-bi equation.
Namely, it was suggested to decompose the effective action into the 
contributions from over and from below the normalization
point $\mu$.
The problem with this idea is that generically, even for
the {\it strongly} complete partition functions, these
two contributions are of different nature \cite{GMS}:
the one from above $\mu$ can indeed be considered as effective
action, while the one from below $\mu$ is rather
a differential operator acting on functions on the moduli space of
coupling constants. A special procedure is needed to extract
a kind of a kernel from this operator, which can be interpreted
as something similar to an effective action. This can be
done in special approximations and with certain reservations.
An obvious example when such decomposition exists, 
is provided by the
quasiclassical approximation, when the shift
of effective action with the changing $\mu$ 
is described by the tree (one-particle-reducible) diagrams,
but it is not fully satisfactory: there are no such diagrams
in the case of matrix models -- which seems to be a prototype
of the actually interesting situations. 
Despite quasiclassical approximation does not work in this
simple way, it is known \cite{MMMM} that the
relevant shift does occur in the matrix models,
and is indeed similar in many respects \cite{MT} 
to the ansatz of \cite{Ver},
but no straightforward way is known to derive it ``from
the general principles'' in the leading $N^{-1}$ approximation.
Worse than that, the very notion of normalization point
becomes somewhat subtle for the {\it weakly} complete
partition functions, of which the matrix models is
an example.

In this note we do not suggest any definite conclusion from
this description of problems.

As already mentioned, it can happen
that the controversy is resolved if the full power of
integrability (the full set of the Ward identities for the
weakly complete partition function) is taken into account.
Clarification of this option is intimately related to understanding the origins
of a generalized AdS/CFT correspondence, i.e. of the representation of generic
exact partition functions (not necessarily of a CFT) in terms of gravity
theories (not necessarily on AdS). The full set of Ward identities for the
weakly complete partition functions is implied by invariance of the integral
(\ref{pf}) under arbitrary coordinate transformations in ${\cal A}$ (arbitrary
changes of integration variables) \cite{MM}. In the strongly complete case,
when the Ward identities are linear differential equations and each of them
generates an RG flow, Polchinski's flow (\ref{PWI}) being just one of them,
the general covariance in the ${\cal A}$ space leads to the general covariance
in the ${\cal M}$ space. This means that the partition function is essentially
an invariant of general coordinate transformations in ${\cal M}$, thus giving
a most natural object for a gravity theory on ${\cal M}$. This property is
broken in realistic models by boundary conditions $\varphi_0$ and by the lack
of the strong completeness: only the weak one is usually natural. Sometimes,
the deviation from linearity in the Ward identities can be interpreted as "a
quantum effect" in (\ref{pf}): the terms non-linear in $t$-derivatives come
from the change of measure, not action, in (\ref{pf}) under reparametrizations
of ${\cal A}$ \cite{MM}, but this does not make one free of handling these
non-linearities. It is very natural to assume that in the weakly complete case
$Z(t)$ is still interpretable in terms of a gravity theory on ${\cal M}$, but
is a slightly less trivial object than just an invariant. Within the AdS/CFT
correspondence \cite{AdS/CFT}, this is rather a wave function, similarly to
the considerations from the perspective of integrability theory \cite{GMS}. In
other words, $Z(t)$ is an invariant not of the ${\it Diff} {\cal M}$ subgroup
of ${\it Dop} {\cal M}$ but of some other subgroup which should be a kind of a
smooth deformation of ${\it Diff} {\cal M}$. The relation between the Feynman
diagrams in field theory and the generators of ${\it Dop} {\cal M}$ \cite{GMS}
implied by considerations of \cite{CK} is one of the new tools to attack the
problem.

Another, less attractive but simpler option is to
sacrifice the {\it exact} RG (i.e. abandon the hope to find 
{\it some} substitute  of Polchinski's equation, which indeed holds 
exactly in the weakly complete case and still has something
to do with the diffeomorphism group ${\it Diff} {\cal M}$),
and try to resolve the controversy near the critical points,
in neglect of non-singular contributions. Then the power of the RG
methods will be strongly reduced (just to the level they have
in conventional quantum field theory), and even if it can help,
this is hardly a satisfactory result. A slightly more
interesting version of this option is that the {\it linear} RG
is an effective object, valid for description of effective actions
near the critical points, but different near different points.
When extrapolated from the vicinity of one point to another,
RG dynamics becomes non-linear, and linear RG equations are
nothing but an approximation to non-linear 
Lap\-la\-ce/Vi\-ra\-so\-ro/Hanil\-ton-Ja\-co\-bi
equations (which generically are not even quadratic).
This option, familiar from the studies of matrix models
\cite{MM,Mamorev,MMMM}
can be the closest in spirit to  the suggestion
of refs.\cite{Ver}.

The rest of this paper contains just a brief comment on the
terminology, used in above considerations.

\section{Partition functions}

The partition function 
 
\be
Z(G;\varphi_0;t) = 
\int_{{\cal A};\varphi_0} D\phi 
\exp \left( -\frac{1}{2}\phi G\phi + {A}(t;\phi)\right)
\equiv \la 1 \ra
\label{pf}
\ee
depends on:
\begin{itemize}
\item the background fields $\varphi_0$;

\item the coupling constants $t$;

\item the metric $G$.
\end{itemize}

\subsection{The fields}

In eq.(\ref{pf}) ${\cal A}$ denotes the space of quantum fields 
(domain of integration in the functional integral).
In $D$-dimensional field theory 
it is a $D$-loop space of maps from the $D$-dimensional
``world sheet'' (space-time) $W$ into a target space
$T$:  $\ \ {\cal A} = \{\hbox{maps}\ W \rightarrow T\} =
L_W^D(T)$. In the (second quantized) string field theory
 $W$ is itself a space
of loops in the space-time 
(while in the first-quantized theories
the space-time plays instead the role of the target space $T$).
When $W$ is not compact, one needs to impose the
boundary conditions at its boundary:
$\varphi_0 \in \{\hbox{maps}\ \partial W \rightarrow T\}$.
In most cases partition functions are non-vanishing
only when the boundary conditions belong to some (co)homologies
of the target space, $\varphi_0 \in H^*(T)$.

\subsection{The coupling constants \label{comp}}

The coupling constants parametrize the shape of the action

\be
A(t,\phi) = \sum_{n \in B} \tn {\cal O}_n(\phi)
\ee 
where the sum goes
over some complete set $B$ of functions ${\cal O}_n(\phi)$,
not obligatory finite or even discrete. The space ${\cal M}
\subset \hbox{Fun}({\cal A})$ 
of actions, parametrized by the coupling constants $\tn$, 
is refered to as the moduli space of theories. The actions
usually take values in numbers or, more generally, in certain
rings, perhaps, non-commutative. The space 
$\hbox{Fun}({\cal A})$ of all functions of $\phi$ is always a
ring, but this need not be true about the moduli space ${\cal M}$,
which could be as small a subset as one likes. However,
the interesting notion of partition function arises only if
the completeness requirement is imposed on ${\cal M}$ \cite{UFN2}.
There are two different degrees of completeness, relevant
for discussions of partition functions. In the first case 
(strong completeness) the
functions ${\cal O}_n(\phi)$ form a {\it linear} basis in 
$\hbox{Fun}({\cal A})$, then ${\cal M}$ is essentially the same
as $\hbox{Fun}({\cal A})$ itself.
In the second case (weak completeness) the functions
${\cal O}_n$ generate $\hbox{Fun}({\cal A})$ as a ring, i.e.
arbitrary function of $\phi$ can be decomposed into a sum of
{\it multi}linear combination of ${\cal O}_n$'s. 
In the case of strong completeness the notion of RG is absolutely
straightforward \cite{Pol,RGrev,GMS}, but 
there is no clear idea how RG can be formulated
in the case of weak completeness
(which is more relevant for most modern considerations\footnote{
The difference between the strong and weak completeness
was recently rediscovered \cite{Mi} 
(for an earlier related analisys see \cite{Br}) in attempts
to test the relation between Polchinski's and the
holomorphic RG with the help of the standard formalism of matrix
models \cite{MM,Mamorev,MMMM}. 
Also, to avoid confusion let us emphasize that 
in the case of $D$-dimensional field theories
the set of functions
$\hbox{Fun}({\cal A})$ includes not just polynomials of $\phi$,
but also the derivatives of $\phi$ along various directions in $W$.
}). 

\subsection{The metric}

In quantum field theory the metric $G$ is needed at least for
two purposes:
to define perturbation theory as a formal sum over
Feynman diagrams and to regularize the original functional
integral.
Regularization is needed whenever $\hbox{Vol} {\cal A} =
\int_{{\cal A}} D\phi = \infty$, then the factor
$\exp \left(-\frac{1}{2}\phi G \phi\right)$ helps to make
integrals finite (see below).

One often explicitly extracts from $A(\phi)$ not only the quadratic term
$-\frac{1}{2}\phi G\phi$ with the metric, but also the linear source term
$J\phi$ and the ``vacuum energy'' $A_0 = A(\phi = 0)$.

\subsection{Normalization point}

The Kadanoff-Wilson RG occurs when a filtration is
defined in the space of fields, i.e. a map from
positive numbers (real or integer, accordingly the RG
is continuous or discrete) into the set $2^{{\cal A}}$
of the subsets of ${\cal A}$, such that

\be
{\cal A}_{\infty}\subset {\cal A}_{\lambda} \subset {\cal A}_\mu \subset
{\cal A}_0 ={\cal A}/H^*(T)
\ \ \ \forall \mu < \lambda 
\ee
Accordingly the complements ${\cal B}_\mu =
{\cal A}/{\cal A}_\mu$ of ${\cal A}_\mu$ in ${\cal A}$
satisfy

\be
{\cal B}_0 \subset {\cal B}_\mu \subset
{\cal B}_\lambda \subset {\cal B}_\infty = {\cal A}
\ \ \ \forall \mu < \lambda 
\ee
One usually assumes that in the infra-red limit,
$\mu = 0$, the space ${\cal B}_\mu$ shrinks to the space
${\cal B}_0=H^*(T)$ of vacua, while in the ultraviolet limit,
$\mu = \infty$, the space ${\cal B}_{\infty}$ coincides with
the entire ${\cal A}$.

Given such a filtration, one can define a $\mu$-dependent
partition function

\be
Z_\mu(G;\varphi_\mu;t) =
\exp \left( -\frac{1}{2}\varphi_\mu G\varphi_\mu\right) 
\int_{{\cal A}_\mu;\varphi_\mu} D\phi 
\exp \left( -\frac{1}{2}\phi G\phi + 
{A}(t;\phi + \varphi_\mu)\right)
\equiv \exp \left( -\frac{1}{2}\varphi_\mu G\varphi_\mu
+ A_\mu(t;\varphi_\mu)\right) 
\label{pfmu}
\ee
The background fields $\varphi_\mu \in {\cal B}_\mu$
and the functional integral goes over ${\cal A}_\mu$.
We assumed that the metric does not mix the fields from
${\cal A}_\mu$ and ${\cal B}_\mu$, but the other terms in the
action unavoidably do. 

If the set of functions ${\cal O}_n$ is {\it strongly} complete, 
then we are dealing with a renormalizable family
of field theories, and the new action $A_\mu(t)$ belongs
to the same moduli space ${\cal M}$, i.e. such $t_\mu(t)$
exist that

\be
A_\mu(t;\phi) = A(t_\mu,\phi)
\ee
and 

\be
Z_\mu(t;\phi_\mu) = Z_\infty(t_\mu,\phi_\mu)
\ee
Then for two normalization points $\mu < \lambda$ we have:

\be
Z_\mu(t;\varphi_\mu) = 
\int_{{\cal A}_\mu/{\cal A}_\lambda = {\cal B}_\lambda/{\cal B_\mu}}
Z_\lambda(t;\phi + \varphi_\mu) D\phi
\ee
This procedure is known as Kadanoff transformation.

\section{Polchinski's exact RG equation}

A simple way to vary the normalization point $\mu$ is provided by
the change of metric \cite{Pol}. It is enough to 
introduce a $\mu$-dependent family of metrics $G_\mu$, such that
${\cal A}_\lambda = \hbox{supp}\ (G_\mu^{-1})$. This motivates the
study of metric dependence of partition function.

The variation of $Z(t)$ with the variation of metric $G$ is:

\be
\delta Z(t) =  \la \phi \delta G \phi \ra
\ee
A Ward identity \cite{Pol}\footnote{
It follows from the obvious identity
$$
0 = \int D\phi \frac{\partial}{\partial\phi}\left\{
\delta G^{-1}\left(\frac{\partial}{\partial \phi} + 2 G\phi\right)
\exp\left(-\frac{1}{2}\phi G\phi + A(\phi)\right)
\right\}
$$
}
states that

\be
\la \phi \delta G \phi \ra = -
\la \delta G^{-1}\left(\frac{\partial A}{\partial \phi}
\frac{\partial A}{\partial \phi} +
\frac{\partial^2 A}{\partial \phi^2} + G\right)\ra
\label{PWI}
\ee
For concrete actions of particular models 
(which do not satisfy completeness requirement), the r.h.s.
is an average of a new operator and is not expressible through
$Z(t)$. However, for the partition function, built with the help
of complete sets of functions the situation is different. 
For the {\it strongly} complete set of ${\cal O}_n$ 
one can simply define $\delta \tn$ from

\be
\delta A(\phi)  =
\delta G^{-1}\left(\frac{\partial A}{\partial \phi}
\frac{\partial A}{\partial \phi} +
\frac{\partial^2 A}{\partial \phi^2} + G\right)
= \sum_n \delta \tn {\cal O}_n(\phi)
\label{shift}
\ee
without any averaging.
This is the situation described in terms of RG \cite{Pol}. Then, eq.(\ref{PWI})
can be rewritten as a {\it linear} differential equation

\be
\delta Z(t) =  \la \phi \delta G \phi \ra =
\sum_n \delta \tn \frac{\partial Z(t)}{\partial \tn}
\equiv \delta G\cdot \hat v(t) Z(t)
\label{PWIs}
\ee
In a {\it weakly} complete case one can not define $\delta \tn$
from (\ref{shift}), but an analogue
of (\ref{PWIs}) still exists, only it involves a
differential operator $\hat\Delta$, not obligatory linear in
$t$-derivatives:

\be
\delta Z(t) =  \la \phi \delta G \phi \ra =
-\delta G\cdot \hat\Delta(t) Z(t)
\label{PWIw}
\ee
or

\be
\hat L(t) Z(t) \equiv
\left(\frac{\partial}{\partial t^{(2)}} +
\hat\Delta(t) \right) Z(t) = 0
\label{Vir}
\ee
In the case of the matrix models this is exactly (one of the)
$W$- or Virasoro constraints \cite{MM,Wc,Mamorev}. Of course, there
are many more Ward identities for complete partition
functions, besides (\ref{PWI}). Some of them are
actually linear in $t$-derivatives, but in the weakly
complete cases these linear equations do not contain
$G$-derivatives ($N$-variations in the case of matrix
models), and do not help, at least
in any straightforward way, to formulate an RG equation.

\section{Quasiclassical RG}

As an example of Kadanoff-Polchinski's procedure consider 
a theory with $N$ copies of every field $\phi$ and 
partition function

\be
Z_N(t) = \int
\prod_{i=1}^N D\phi^i \exp \ {\hbar}^{-1}
\left(-\frac{1}{2} \phi^i G_{ij}\phi^j +
\sum_n \tn {\cal O}_n(\phi)\right) 
\ee
Let us now add one more field $\phi$. Such
changing of $N$ can be considered as a result of a change
of metric: switching on or off some components of $G^{ij}$.
This change $N \rightarrow N+1$ 
is not infinitesimal, but in the quasiclassical
approximation, i.e. in the leading order in $\hbar$ expansions,
a result similar to (\ref{PWIw}) holds:

\be
Z_{N+1}(t) = \int D\phi e^{-\frac{1}{2\hbar} G\phi^2}
\int \prod_{i=1}^N D\varphi^i \exp\ \hbar^{-1}
\left(-\frac{1}{2\hbar}\varphi^i G_{ij}\varphi^j +
\sum_n \tn {\cal O}_n(\phi + \varphi)\right) \times \\ \times
\exp \ {\hbar}^{-1}\left(\phi\sum_n \tn
\frac{\partial {\cal O}_n}{\partial \phi}(\varphi)
+\frac{1}{2}\phi^2 \sum_n \tn
\frac{\partial^2 {\cal O}_n}{\partial \phi\partial\phi}(\varphi)
+ O(\phi^3) \right) =\\ =
\left.\la \exp \ -\frac{1}{2\hbar\tilde G} \left( \sum_{m,n}
\tm\tn \frac{\partial {\cal O}_m}{\partial \phi}
\frac{\partial {\cal O}_n}{\partial \phi}
+ O(\hbar)\right) \ra\right._{\hspace{-0.3cm}N}\equiv e^{\hat D_N(t)}Z_N(t)
\ee
with $\tilde G = G + \sum_n \tn 
\frac{\partial^2 {\cal O}_n}{\partial \phi\partial\phi}$.
We assumed here that $\phi = \phi^{N+1}$ enters the original
Lagrangian in the same way that all the other $\phi$'s,
e.g. all the operators ${\cal O}_n$ are $U(N+1)$ symmetric,
and took $G_{i,N+1}=0$ for symplicity.
This tree-like formula
manifests the relation between Kadanoff-Wilson RG\footnote{
In theories with the power-like scaling laws one often
supplements the ``integration-out'' procedure by
afterall rescaling of the world sheet $W$ ``back to its original 
size''. Though important
for some interpretations of RG equations, this additional
procedure is not essential for our purposes.
}
in the quasiclassical  approximation and Polchinski's RG eqs.
It describes the shift of the classical action provided
by one-particle-reducible diagrams.

In matrix models this integration-out procedure,
changing the size of $N\times N$ matrices does not lead to such
a shift (unless there are $U(1)$ factors in the symmetry group). 
The difference is that instead of eliminating a
single $\phi = \phi^{N+1}$, in the case of matrix model
  one needs to eliminate the 
whole vector $\phi^i = \phi^{i,N+1}$ from Hermitean matrix
$\phi^{ij}$, and the relevant operators ${\cal O}_n(\phi)$
(like $Tr \phi^n$) are bilinear in $\phi^i$. Therefore
there are no contributions of the order $\hbar^{-1}$,
and quasiclassical approximation is not a relevant
approximation in the case of matrix models. Its proper substitute
is the $N^{-1}$ expansion, where a variety of possibilties
occurs, depending on the assumed $N$-dependence of coupling
constants. 

\section{Description in terms of the diffeomorphisms}

The adequate description of RG for the {\it strongly} complete
partition functions is in terms of  diffeomorphisms \cite{GMS}:

\be
Z(G'; t) = Z(G; t'(t)) = e^{\hat V(t)} Z(G; t)
\label{RGe}
\ee
Here $\hat V(t;G',G) = \sum V_n(t)\frac{\partial}{\partial\tn}$
is a vector field, so that $e^{\hat V(t)}$
is a differential operator.
Its relation to the shift $t'^{(n)} - \tn = \tilde V_n(t)$
is provided by generic identification \cite{GMS}
of elements from ${\it Diff}{\cal M}$ and ${\it Shift}{\cal M}$
groups acting on the moduli space of coupling constants

\be
\exp({\hat V}) = \ :\ \exp(\hat{\tilde V})\ :
\ee
The RG equation (\ref{RGe}) represents original (bare)
partition function as an action of an {\it operator} on the new
(renormalized) partition function. Generically, the vector
field $\hat{\tilde V}$
decomposes into a $\partial/\partial t^{(0)}$-piece and
all the other $t$-derivatives. The first piece can be
considered as generating an additive correction to the effective
action $S = \log Z$, while the remaining part of $\hat V$
generates shifts of the other couplings. In other words,
one can represent (\ref{RGe}) in a different form:

\be
S(G'; t) = S_0(t) + S(G; t'(t))
\label{decom}
\ee
where $S$ is supposed to depend only on $\tn$ with $n > 0$,
and 

\be
S_0(t) \equiv t^{(0)} - t'^{(0)}(t) = 
\log \left( \exp({\hat V(t)}) 
e^{t^{(0)}}\right)
\ee
Relation (\ref{decom}) describes a decomposition of the type
suggested in \cite{Ver}. Moreover, like requested in \cite{Ver},
the two items at the r.h.s. of (\ref{decom}) satisfy the
closely related equations. Indeed, a pair of relations, 

\be
\hat L(t) Z(G';t) = 0 , \nn \\
\hat L(t) e^{t^{(0)}} = \ const\cdot e^{t^{(0)}}
\label{orig}
\ee
where the first one is the Ward identity (\ref{Vir}) and the
second one reflects the fact that the trivial
partition function $\exp \left(t^{(0)}\right)$ is usually an
eigenstate of $\hat L(t)$, 
turns into:

\be
\hat{\cal L}(t) Z(G;t) = 0, \nn \\
\hat{\cal L}(t) \exp S_0(t) = \ const\cdot \exp S_0(t)
\label{eqns}
\ee
with a new operator

\be
\hat{\cal L}(t) \equiv \exp({\hat V(t)}) \hat L(t)
\exp(-{\hat V(t)}) 
\ee
The second relation in (\ref{eqns}) implies that $S_0(t)$
is non-vanishing when the eigenvalue in (\ref{orig})
is different from zero.
If rewritten in terms of effective actions $S(t)$
and in the quasiclassical limit, when $\partial^n Z \rightarrow
Z(\partial S)^n$,
the equations (\ref{eqns})
acquire the form of the Hamilton-Jacobi equations. They are
quadratic in $S(t)$ if the differential operator in (\ref{PWIw})
is of the second order in $\partial/\partial t$, what is often
the case in some simple models.

Unfortunately, the above reasoning is mixing two different
things, which are not obligatory compatible: the RG
equation (\ref{RGe}), occuring for the {\it strongly}
complete partition functions, and the Laplace-like
equation (\ref{Vir}), requiring only {\it weak} completeness.
In the strongly complete case, the non-linear (in coupling
derivatives) equation, even if occurs, can be always rewritten
as a linear equation. In fact, one can easily make a
weakly complete model strongly complete, by adding all the
newly emerging operators to the action $A(t;\phi)$, then,
if the product ${\cal O}_m{\cal O}_n$ is added with the coefficient
$t^{(m,n)}$, we have an identity
$\partial^2Z / \partial t^{(m)}\partial t^{(n)} =
{\partial Z}/{\partial t^{(m,n)}}$.
Alternatively, in the weakly complete case one could try to
interpret (\ref{PWIs}) as a substitute for RG equation, but
then in (\ref{RGe}) we get $\ \exp (\hat V(t))=P\exp(\int\delta G\cdot\hat
v(t))$ substituted by $\exp \hat D = P\exp (\int \delta G\cdot\hat\Delta(t))$,
which is an element of ${\it Dop} {\cal M}$, but no longer of ${\it Diff} {\cal
M}$.

\section{Acknowledgements}

We are indebted for discussions to E.Akhmedov, R.Brustein,
 A.Gerasimov, A.Gorsky, A.Losev, A.Marshakov and T.Tomaras.
We acknowledge the hospitality and support of the ESI, Vienna,
where this paper was written. Our work is partly supported
by the grants: RFBR 98-01-00328,
INTAS 99-0590 (A.Mir.), RFBR 98-02-16575, the Russian
President's Grant 00-15-99296 (A.Mor.) and by CRDF grant \#6531.


\begin{thebibliography}{12}


\bibitem{RG} N.Bogolubov and D.Shirkov, {\sl Introduction to the Theory of
Quantum Fields}, Moscow, 1957;\\
M.Gell-Mann and F.E.Low, Phys.Rev.,  {\bf 95} (1954) 1300;\\
L.P.Kadanoff, In: {\sl  Varenna 1970, Proceedings, Critical Phenomena}, New
York 1971, 100-117;\\
K.G.Wilson and J.Kogut, Phys.Rept., {\bf 12C} (1974) 77;\\
A.Polyakov, {\sl Gauge Fields and Strings}, Harwood Academic Publishers, 1987.
\bibitem{Pol} J.Polchinski, Nucl.Phys., {\bf B231} (1984) 269.
\bibitem{RGrev} For the text-books on RG see: \\
A.Z.Patashinsky and V.N.Pokrovsky, {\sl Fluctuation Theory of Phase
Transitions}, Oxford, England: Pergamon Pr., 1979 (International Series In
Natural Philosophy, Vol.98);\\
S.K.Ma, Rev.Mod.Phys., {\bf 45} (1973) 589;\\
and also recent reviews:\\
T.R.Morris, Prog.Theor.Phys.Suppl. {\bf 131} (1998) 395, hep-th/9802039;\\
C. Bagnuls and C. Bervillier, hep-th/0002034.
\bibitem{Kri} I.Krichever, Comm.Pure Appl.Math., {\bf 47} (1994) 437;\\
B.Dubrovin, Nucl.Phys., {\bf B379} (1992) 627; hep-th/9407018;\\
H.Itoyama and A.Morozov, Nucl.Phys., {\bf B491} (1997) 529, hep-th/9512161
\bibitem{GKMMM}
A.Gorsky, I.Krichever, A.Marshakov, A.Mironov and A.Morozov,
Phys.Lett., {\bf B355} (1995) 466, hep-th/9505035;\\
A.Gorsky, A.Marshakov, A.Mironov and A.Morozov,
Nucl.Phys.,  {\bf B527} (1998) 690, hep-th/9802007.
\bibitem{AdS/CFT} J.Maldacena, Adv.Theor.Math.Phys., {\bf 2} (1998) 231,
hep-th/9711200;\\
S.Gubser, I.Klebanov and A.Polyakov, Phys.Lett., {\bf B428} (1998) 105,
hep-th/9802109;\\
E.Witten, Adv.Theor.Math.Phys., {\bf 2} (1998) 253, hep-th/9802150.
\bibitem{Hol} A.Gerasimov, {\it unpublished}; \\
E.Akhmedov, Phys.Lett., {\bf B442} (1998) 152, hep-th/9806217\\
E.Alvarez and C.G\'omez, Nucl.Phys., {\bf B541} (1999) 441, hep-th/9807226.
\bibitem{Holl} D.Z.Freedman, S.S.Gubser, K.Pilch and N.P.Warner,
hep-th/9906164;\\
L.Girardello, M.Petrini, M.Porrati and A.Zafaroni, Nucl.Phys., {\bf B569}
(2000) 451, hep-th/9906164; JHEP, {\bf 9812} (1998) 022, hep-th/9810126;\\
V.Balasubramanian and P.Kraus, Phys.Rev.Lett., {\bf 83} (1999) 3605;
hep-th/9903190;\\
K.Skenderis and P.K.Townsend,  Phys.Lett., {\bf B468} (1999) 46-51,
hep-th/9909070;\\ O.DeWolfe, D.Z.Freedman, S.S.Gubser and A.Karch,
hep-th/9909134.
\bibitem{Polyakov} A.Polyakov, Proceedings of the Les Houches School, 1993;\\
A.Polyakov, Nucl.Phys.Proc.Suppl., {\bf 68} (1998) 1; hep-th/9711002.
\bibitem{Ver} J.de Boer, E.Verline and H.Verlinde, hep-th/9912012;\\
E.Verlinde and H.Verlinde, hep-th/9912018;\\
E.Verlinde, Class.Quant.Grav. 17 (2000) 1277-1285, hep-th/9912058;\\
J.Khoury and H.Verlinde, hep-th/0001056.
\bibitem{GMS} A.Gerasimov, A.Morozov and K.Selivanov, hep-th/0005053.
\bibitem{UFN2} A.Morozov, Sov.Physics Uspekhi, {\bf 35} (1992) 671.
\bibitem{Vir} M.Fukuma, H.Kawai and R.Nakayama, Int.J.Mod.Phys., {\bf A6}
(1991) 1385;\\ R.Dijkgraaf, E.Verlinde and H.Verlinde, Nucl.Phys., {\bf B348}
(1991) 435. \bibitem{MM} A.Mironov and A.Morozov, Phys.Lett., {\bf B252} (1990)
47-52;\\ J.Ambjorn, J.Jurkiewicz and Yu.Makeenko, Phys.Lett., {\bf 251B}
(1990) 517-524;\\
H.Itoyama and Y.Matsuo, Phys.Lett., {\bf 255B} (1991) 202.
\bibitem{Wc}
A.Marshakov, A.Mironov and A.Morozov, Mod.Phys.Lett., {\bf A7} (1992)
1345-1359;\\
C.Ahn, K.Shigemoto, Phys.Lett., {\bf B285} (1992) 42, hep-th/9112057;\\
A.Mikhailov, Int.J.Mod.Phys., {\bf A9} (1994) 873, hep-th/9303129.
\bibitem{Mamorev} A.Morozov, Sov.Physics Uspekhi, {\bf 37} (1994) 1,
hep-th/9303139; hep-th/9502091;\\
A.Mironov, Int.J.Mod.Phys., {\bf A9} (1994) 4355, hep-th/9312212.
\bibitem{MMMM} Yu.Makeenko, A.Marshakov, A.Mironov and A.Morozov,
Nucl.Phys., {\bf B356} (1991) 574-628.
\bibitem{MT} A.Marshakov, A.Mironov and T.Tomaras, {\it in preparation}.
\bibitem{CK} A.Connes and D.Kreimer, Commun.Math.Phys., {\bf 210} (2000)
249, hep-th/9912092; hep-th/0003188.
\bibitem{Mi} M.Li, hep-th/0001193.
\bibitem{Br} R.Brustein and S.P.De Alwis,
``Renormalization group equation for string field theory and matrix models,''
UTTG-08-91, Presented at PASCOS-91 Int. Symp., Boston, MA Mar 25-30,
1991.



\end{thebibliography}
\end{document}